\documentclass{article}
\usepackage[utf8]{inputenc}
\usepackage{xcolor}

\usepackage{amsmath}
\usepackage{graphicx}
\usepackage{natbib}
\setcitestyle{open={(},close={)}}
\usepackage{url} 
\usepackage{hyperref}

\usepackage[]{authblk}

\title{On assessing excess mortality in Germany during the COVID-19 pandemic}
\author[1]{Giacomo De Nicola
\footnote{Corresponding author, \texttt{giacomo.denicola@stat.uni-muenchen.de}
}
}
\author[1]{Göran Kauermann}
\author[2]{Michael Höhle}
\affil[1]{Department of Statistics, LMU Munich, Germany}
\affil[2]{Department of Mathematics, University of Stockholm, Sweden}

\date{}

\begin{document}

\maketitle

\begin{abstract}
Coronavirus disease 2019 (COVID-19) is associated with a very high number of casualties in the general population. Assessing the exact magnitude of this number is a non-trivial problem, as relying only on officially reported COVID-19 associated fatalities runs the risk of incurring in several kinds of biases. One of the ways to approach the issue is to compare overall mortality during the pandemic with expected mortality computed using the observed mortality figures of previous years. In this paper, we build on existing methodology and propose two ways to compute expected as well as excess mortality, namely at the weekly and at the yearly level. Particular focus is put on the role of age, which plays a central part in both COVID-19-associated and overall mortality. We illustrate our methods by making use of age-stratified mortality data from the years 2016 to 2020 in Germany to compute age group-specific excess mortality during the COVID-19 pandemic in 2020.
\end{abstract}

\section{Introduction}
First identified in Wuhan, China, in December 2019, the Coronavirus disease 2019 (COVID-19) caused by the SARS-CoV-2 virus developed into a worldwide pandemic during the spring of 2020  \citep{Velavan-etal:2020}. One of the challenges for scientists has been to evaluate its impact in terms of life loss across different countries and regions of the world. 
A possible way to do this is through directly looking at the number of people who died while they were confirmed to be infected. 
This measure, often defined as COVID-19-associated mortality, is certainly more robust than other pandemic-related quantities such as e.g.\ the number of reported COVID-19 cases, for which it has become clear that there is a non-negligible discrepancy between cases detected through tests and the number of individuals who were infected \citep{lau2021,schneble2021}. Nonetheless, the raw number of COVID-related fatalities can be subject to biases and interpretative issues as well. In particular, this number might also be biased downwards, as COVID-19 cases can still remain unreported until and after the point of death. Moreover,  it is not always straightforward to identify if COVID-19 was the primary cause of death: Some patients might have a SARS-CoV-2 infection, but the actual contribution of the virus to the death might be minimal \citep{vincent2020}. 
To deal with these issues, comparing all-cause mortality is generally considered a more robust alternative for assessing the damage done by the pandemic, and to compare its impact between regions or countries. 
A first look at this matter for Germany was provided by \cite{Stang-etal:2020}, who looked at data from the first wave ranging from calendar weeks 10 to 26 in 2020. The authors came to the conclusion that a moderate excess mortality was observable for this period in Germany, in particular for the elderly. \cite{Morfeld-etal:2021}  consider regional variation in mortality in Germany during the first wave (see also \citealp{Morfeld-etal:2020}). A calculation of the years of life lost over the course of the pandemic in Germany in 2020 was pursued by \cite{rommel2021}. International analyses on excess mortality due to COVID-19 include e.g. \cite{Krieger-etal:2020}, looking at data from Massachusetts, \cite{Vandoros:2020} who focuses on England and Wales, and \cite{Michelozzi:2020} investigating mortality in Italian cities. Global analyses in this direction were pursued by \cite{karlinsky2021} and \cite{aburto2021}.




Monitoring excess mortality has a long tradition as part of analysing the impact of pandemics \citep{johnson2002, simonsen2013}. With the EuroMOMO project, Europe also runs an early-warning system specifically  dedicated to mortality monitoring \citep{mazick2007}. However, no unified methodological definition exists for deciding if the currently observed death counts are higher than what would be expected. 
A very simple approach is to compare the currently observed deaths for a selected time-period with the average of death counts for a similar period in previous years\footnote{https://www.destatis.de/DE/Themen/Gesellschaft-Umwelt/Bevoelkerung/Sterbefaelle-Lebenserwartung/sterbefallzahlen.html}. 
Alternatively, the expected value can be computed by an underlying time-series model based on past values, e.g.\ including seasonality and excluding past phases of excess, as done in the EuroMOMO project (see e.g.\ \citealp{vestergaard2020, norgaard2021}). 
These approaches, however, do not come without problems, as the age structure within a population can change significantly over time. Given that both general and COVID-related mortality are heavily dependent on age (\citealp{Dowd:2020, Levin:2020}), raw comparisons not taking age into account will often lead to biased estimates. 
More sophisticated approaches thus need to adjust for different or changing age structures in the population. The latter point is of particular relevance when looking at aging populations \citep{Kanasi-etal:2016} and the infectious risks for the elderly \citep{Kline:2016}. Such age-adjustments have a long tradition in demography when comparing mortality across different regions with different age-structure \citep{keiding2014, kitagawa1964}. A general discussion on aging populations and mortality can be found in \cite{Crimmins:2019}. 


In this paper, we build on existing methodology to propose two ways of calculating expected mortality taking age into account, respectively at the weekly and at the yearly level. These methods are compared to the existing benchmarks on data from Germany over the years 2016-2019, for which age-stratified information is available. We furthermore apply those methods to assess age group-specific excess mortality in Germany during the COVID-19 pandemic in 2020.  
The remainder of the manuscript is structured as follows. In Section~\ref{sec:exmo_yearly} we look at yearly expected mortality, while the weekly view is pursued in Section~\ref{sec:exmo_weekly}. Section~\ref{sec:discussion} ends the paper with some interpretative caveats and concluding remarks.

\section{Yearly Excess Mortality} \label{sec:exmo_yearly}
We first look at yearly data and tackle the question of whether there was excess mortality in Germany in 2020. 
In order to obtain an age adjustment for mortality data we  calculate expected deaths based on official life tables. Life tables give the probability $q_x$ of a person who has completed $x$ years of age to die before completing their next life-year, i.e.\ before their $x+1^{th}$ birthday. In our analysis we consider the death table provided for the year 2017/2019 from the Federal Statistical Office of Germany \citep{destatis:2020}. The calculation of a life table, as simple as it sounds, is not straightforward and is an age-old actuarial problem. First references date far back, to \cite{Price:1771} and \cite{Dale:1772}. A historical digest of the topic is provided by \cite{Keiding:1987}. Over the last decades, the calculation of the German life-tables made use of  different methods proposed in \cite{Becker:1874}, \cite{Raths:1909} and \cite{Farr:1859}. We will come back to this point and demonstrate that further adjustments are recommendable to relate the expected number of deaths to recently observed ones. 
In particular, with increasing life expectancy, the average age of the German population has been steadily increasing (see e.g.\ \citealp{Buttler:2003}), and this has some effect on the validity of life tables, as discussed in \cite{Dinkel:2002}. Generally, an aging population leads to increasingly high yearly death tolls (see e.g. \citealp{Klenk:2007}). To quantify excess mortality one therefore needs to account for age effects, e.g., leading to the standardized mortality ratio (SMR, see e.g.\ \citealp{rothman2008}). The SMR is defined as the ratio of observed death counts over expected deaths and thus allows for an age adjusted view, meaning that instead of pure death counts one takes the (dynamic) age structure into account.

\begin{figure}
	\centering
	\includegraphics[width=0.5\linewidth]{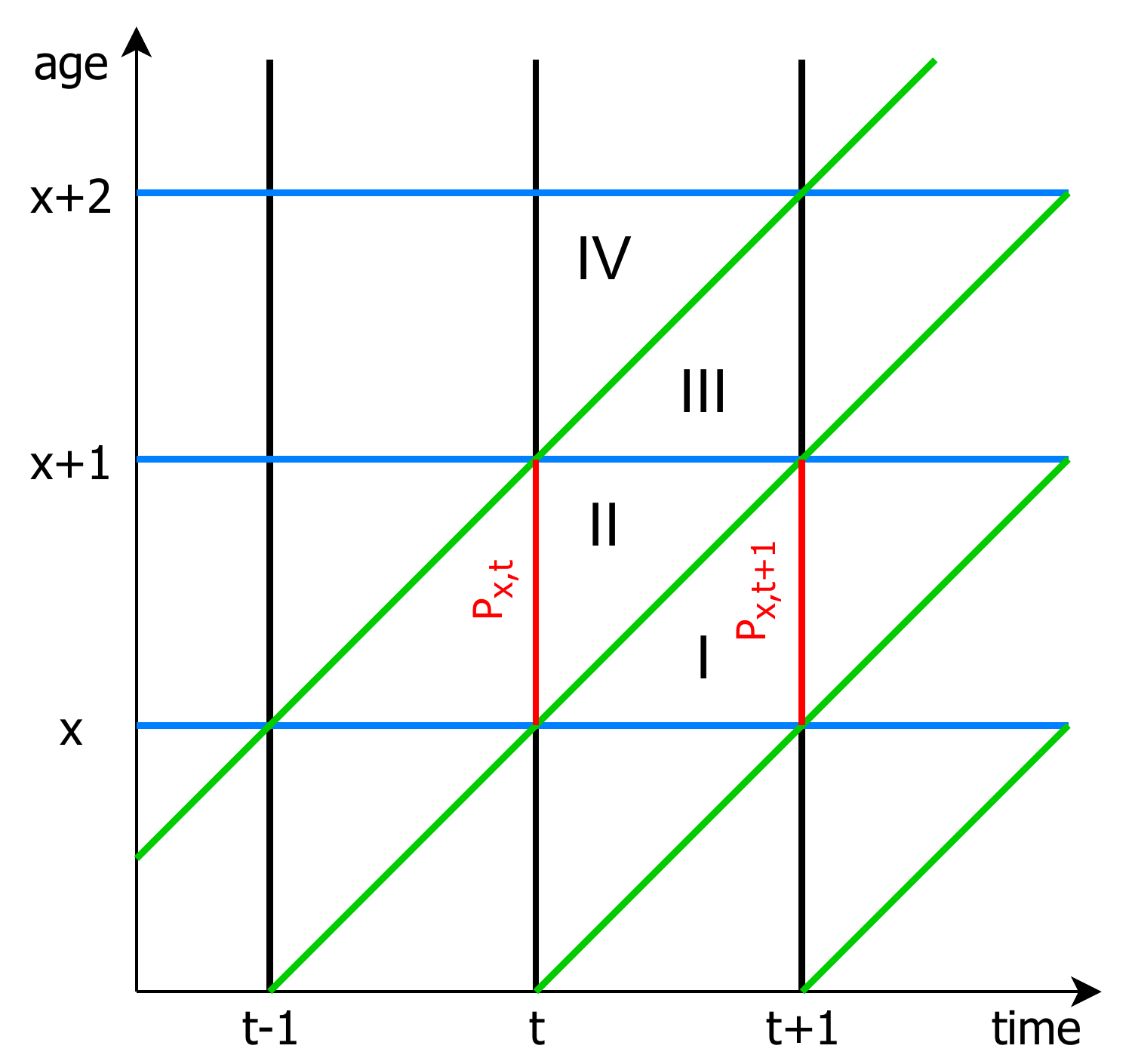}
	\caption{Lexis Diagram}
	\label{Fig:1}
\end{figure}

Calculating excess mortality on a yearly basis requires to calculate expected fatalities using life tables provided by the relevant statistical bureau. We make use of data provided by the Federal Statistical Office of Germany  \citep{destatis:2020}. A straightforward way of obtaining the expected number of deaths for age group $A$ in year $y$ is to calculate 
\begin{align}
	\label{eq:expected1}
	e_{A,y} = \sum_{x \in A} q_x P_{x,y}
\end{align}
where $P_{x,y}$ is the population size of individuals aged $x$ years at the beginning of year $y$, and $q_x$ are, e.g., the age-specific death probabilities in the most recent German life table from the years 2017/19, calculated following \cite{Raths:1909}. More specifically, let $D_x$ be the cumulated number of individuals that died aged $x$ year old, i.e.\ before their $x+1$-th birthday in the considered years 2017 to 2019. Let $P_{x,y} $ denote the population size of $x$ year old individuals on December 31st in year  $y \in \{2016,2017,2018,2019\}$. Then $q_x$ provided in the German life-tables is defined as
\begin{align}
	\label{eq:qx.destatis}
	q_x = \frac{D_x}{\displaystyle \sum_{y=2016}^{2018} \frac{P_{x,y} + P_{x,y+1}}{2} + \frac{D_x}{2}}
\end{align}
We label (\ref{eq:expected1}) in combination with (\ref{eq:qx.destatis}) as {Method 1} below. We will see that this quantity is biased for estimating the expected number of deaths of $x$ year old people in year $y$. To motivate this we look at the Lexis diagram  in Figure \ref{Fig:1}, and for simplicity we replace the calculation in (\ref{eq:qx.destatis})  by looking at a single year only, i.e from $y=t$ to $y=t+1$.  This leads to $D_x = I + II $, where $I$ and $II$ refer to the observed deaths in the two triangles in Figure \ref{Fig:1}. Note that following the calculation principle (\ref{eq:qx.destatis}) of the Statistisches Bundesamt  we would obtain $q_x$ as 
\begin{align}
	\label{eq:expected.destatis1}
	q_x = \frac{D_x}{\displaystyle  \frac{P_{x,t} + P_{x,t+1}}{2} + \frac{D_x}{2}}
\end{align}
where $P_{x,t}$  and $P_{x,t+1}$ are the population sizes of $x$ year old indicated in Figure  \ref{Fig:1}. That is, $q_x$ is the probability of dying in triangles $I$ and $II$. Let us define with  $\tilde{q}_x$ the probability of an individual aged $x$  years old at the beginning of year $t$ (i.e.\ on December 31st in year $t-1$) to die before year $t+1$ starts. In other words $\tilde{q}_x$ is the probability of dying in triangles $II$ and $III$.  In fact, this is the probability we are interested in. It is easy to see that $\tilde{q}_x \ne q_x$.  Assuming that the probability of dying in triangle $I$ is roughly equal to the probability of dying in triangle $II$, and assuming the same relationship for triangles $III$ and $IV$ holds, we can conclude the approximate equivalence
\begin{align}
	\label{eq:qx22}
	\tilde{q}_x = \frac{1}{2} q_x +  \frac{1}{2} q_{x+1}
\end{align}
which  leads to the expected number of deaths
\begin{align}
	\label{eq:expected2}
	\tilde{e}_{A,y} = \sum_{x \in A} \tilde{q}_x P_{x,y}.
\end{align}
We label (\ref{eq:expected2}) as {Methods 2} below. The adjustment is still not complete, and in fact it can be shown that (\ref{eq:expected2}) is biased (see \citealp{Hartz:1983}). Note that individuals dying in triangle $III$ count as $x+1$ years old, so that part of the deaths contributes to a different age group. We may now assume for simplicity that the probability of dying in  triangles $II$ and $III$ is roughly the same, which leads to the following calculation. Let $A = [a_l,a_r]$
\begin{align}
	\label{eq:expected3}
	\hat{e}_{A,y} = 0.5 \cdot \tilde{q}_{a_{l-1}} P_{a_{l-1},y} + \sum_{x = a_l}^{a_{r-1} } \tilde{q}_x p_{x,y} + 0.5 \cdot \tilde{q}_{a_r} P_{a_r,y} 
\end{align}
where $\tilde{q}_{-1} = \tilde{q}_{0}  $ and $P_{-1,y} = P_{0,y}$  gives the approximation for the youngest age group. Accordingly, for $a_r = \max(x) $ we take the full fraction of the last year, that is we add an additional $0.5 \cdot \tilde{q}_{a_r} p_{a_r,y} $ to the formula above. We label (\ref{eq:expected3}) as {Method 3} below. 

\begin{figure}
	\centering
	\includegraphics[width=0.6\linewidth]{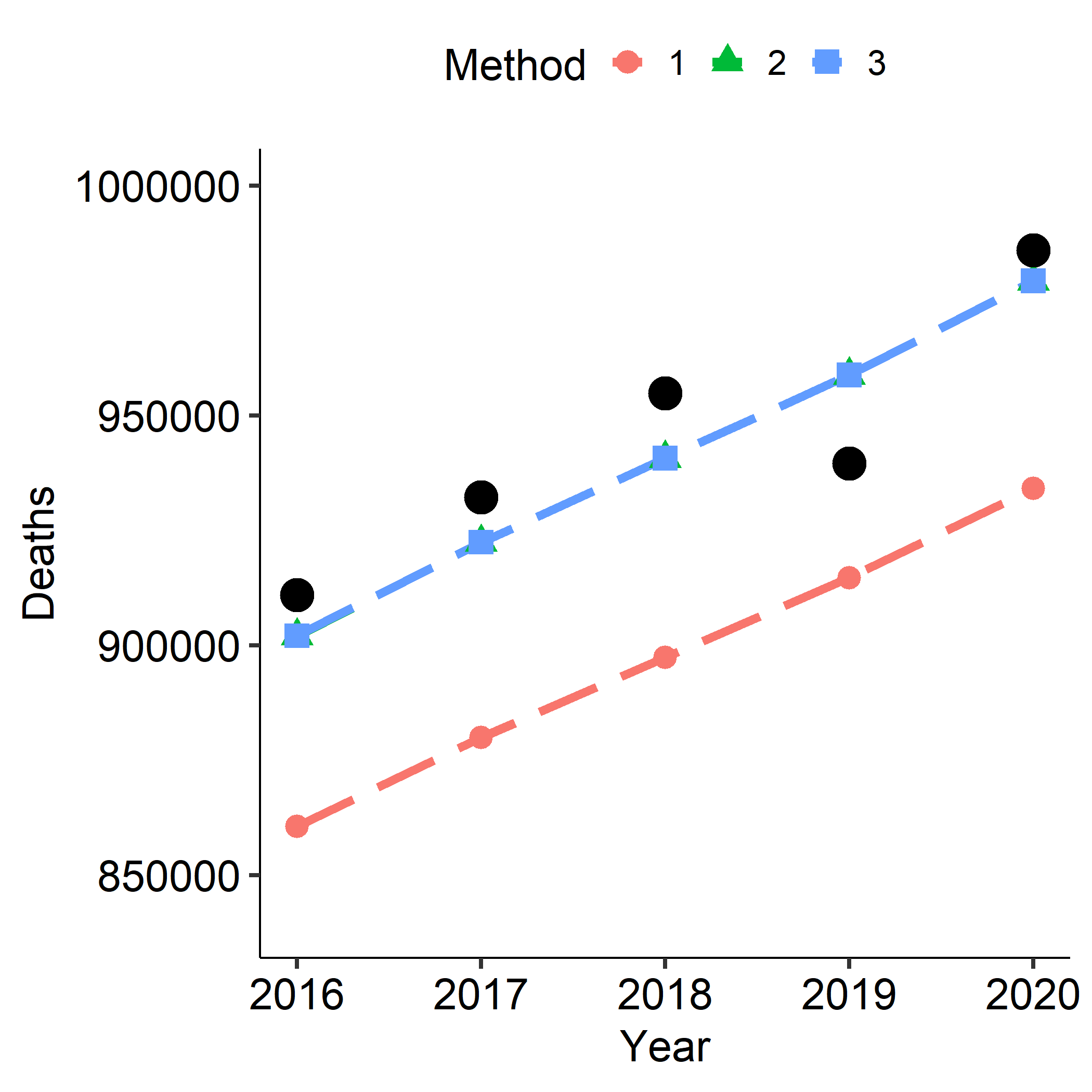}
	\caption{Expected deaths computed by year with the three different methods described, for all age groups combined. Realized fatalities are shown as black dots. Methods 2 and 3 are visually indistinguishable, as age groups are pooled together.}
	\label{fig:all}
\end{figure}

\begin{figure}
	\centering
	\includegraphics[width=0.37\linewidth]{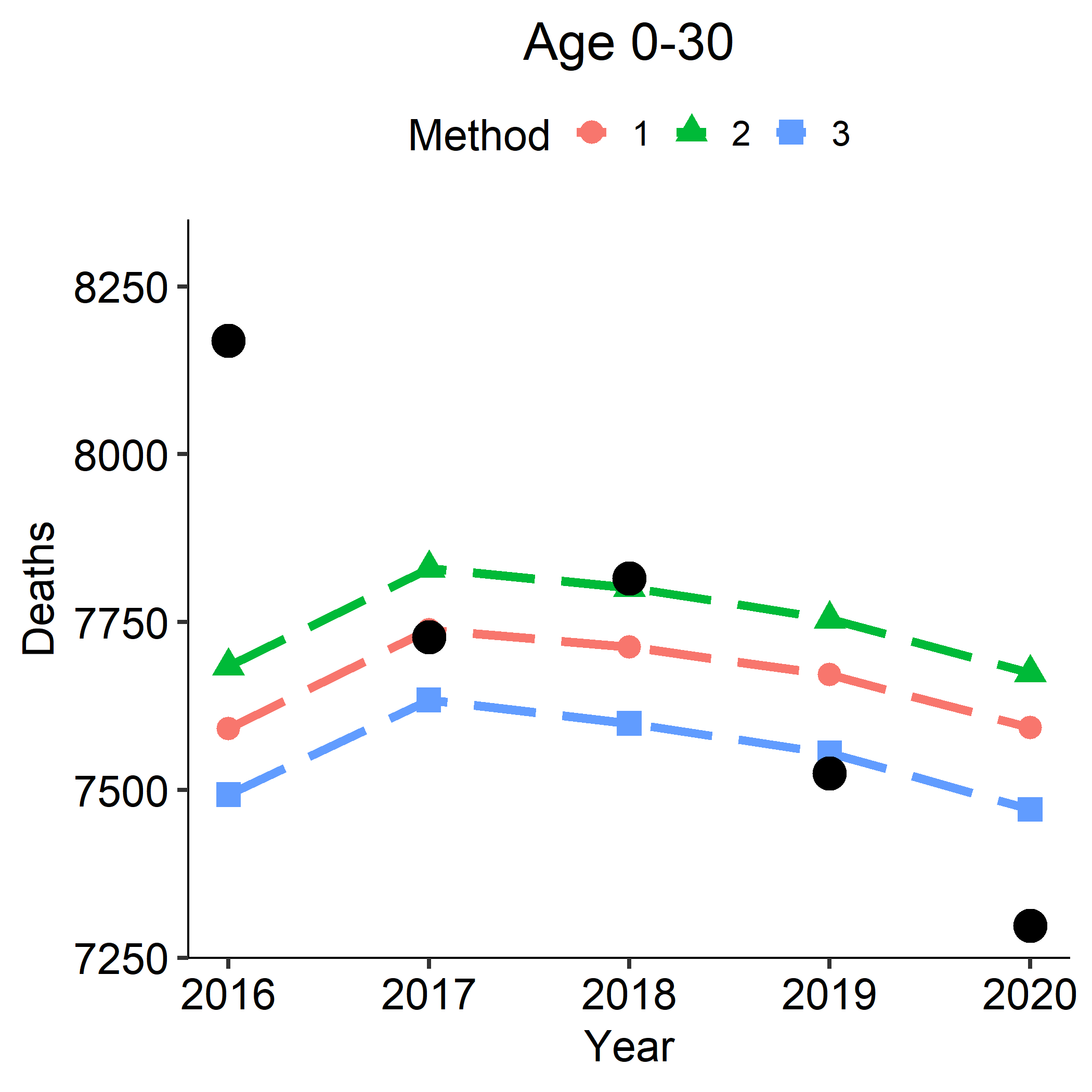}
	\includegraphics[width=0.37\linewidth]{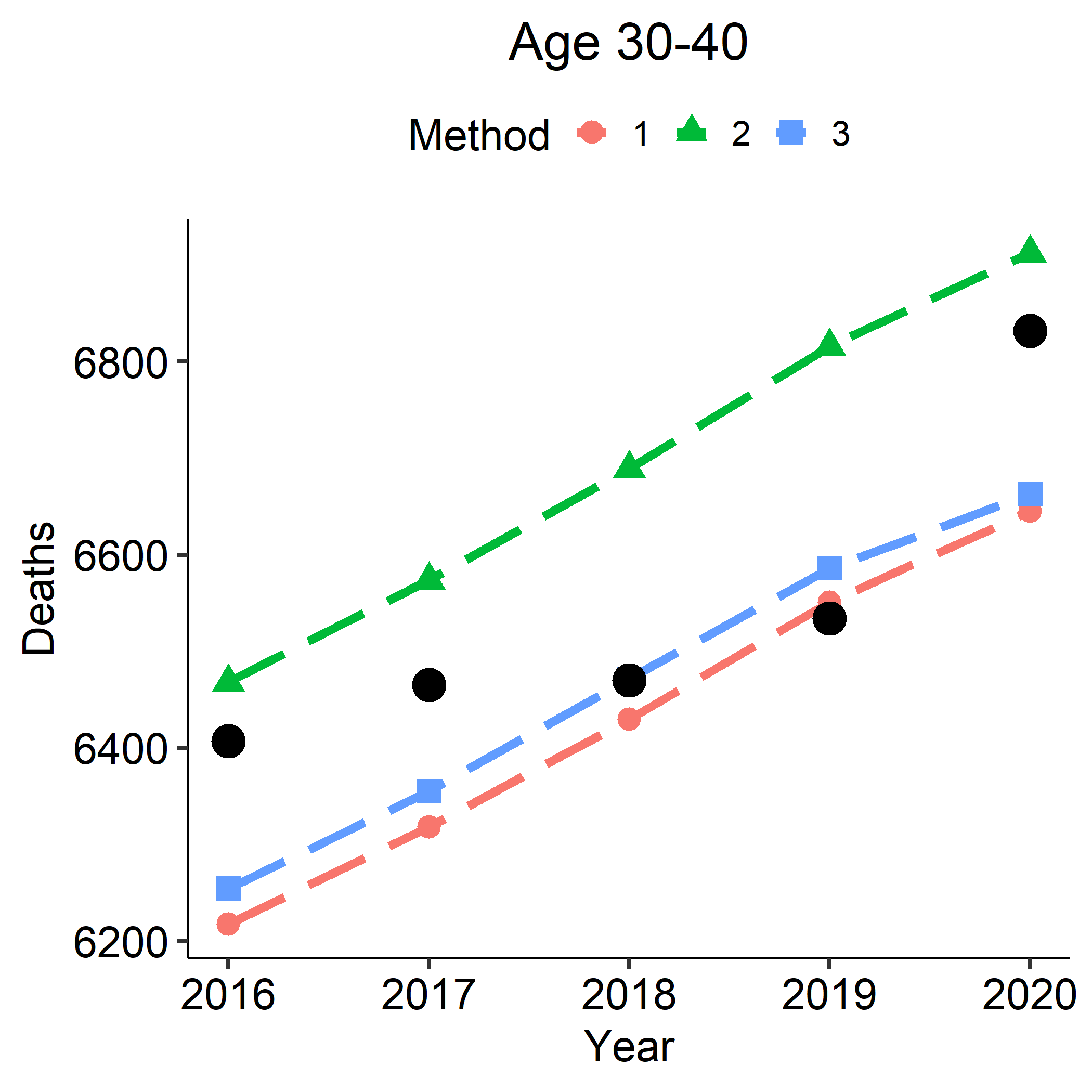}
	\includegraphics[width=0.37\linewidth]{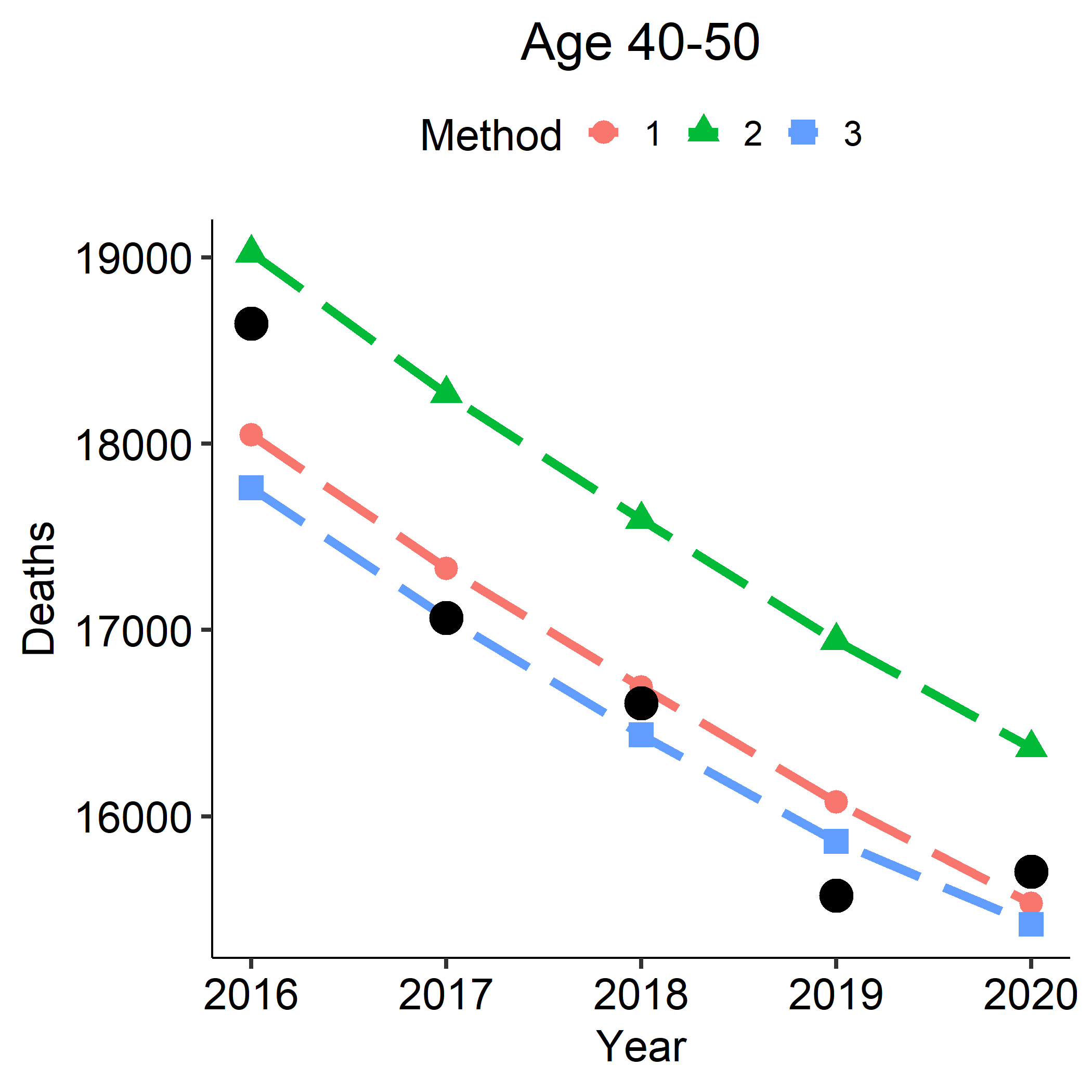}
	\includegraphics[width=0.37\linewidth]{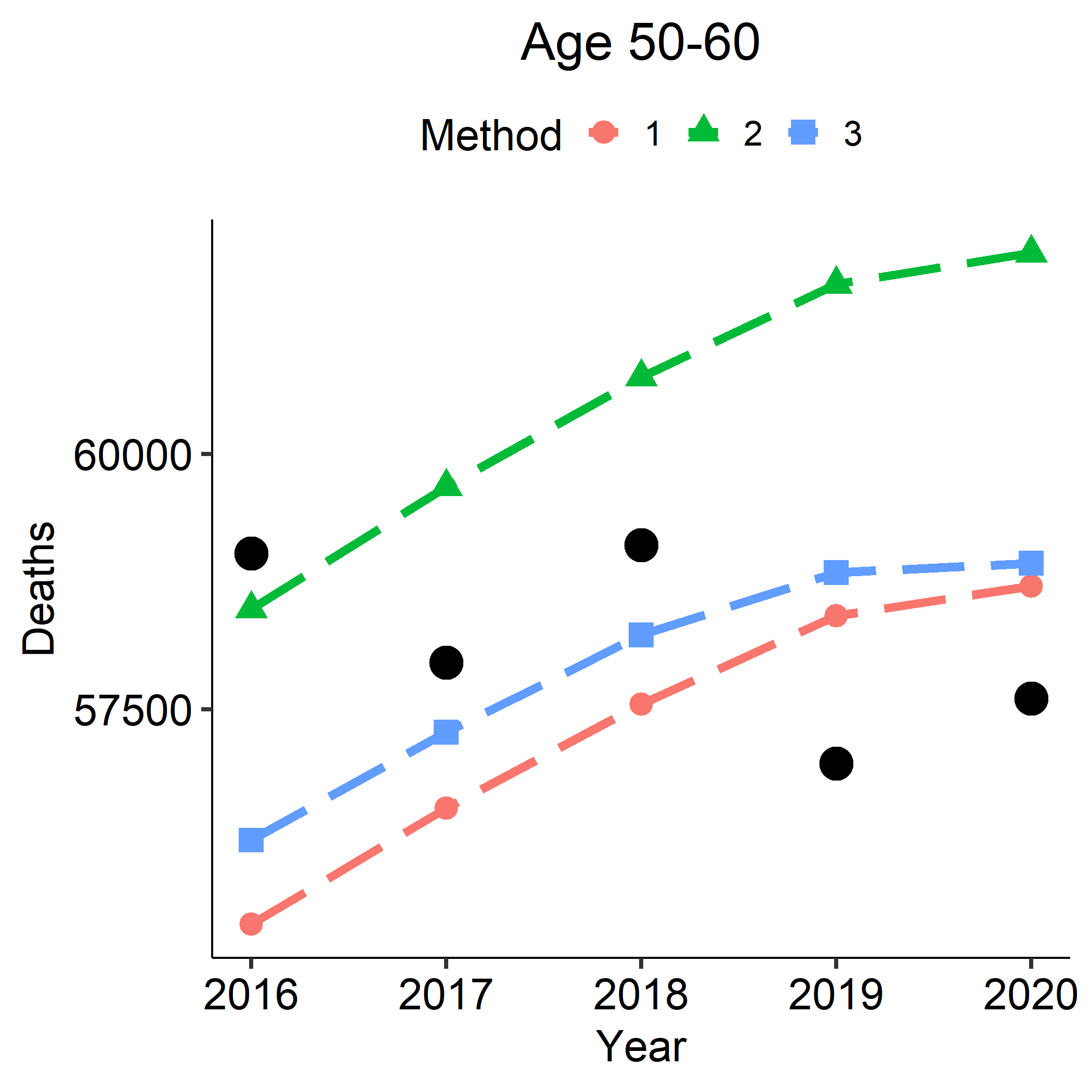}
	\includegraphics[width=0.37\linewidth]{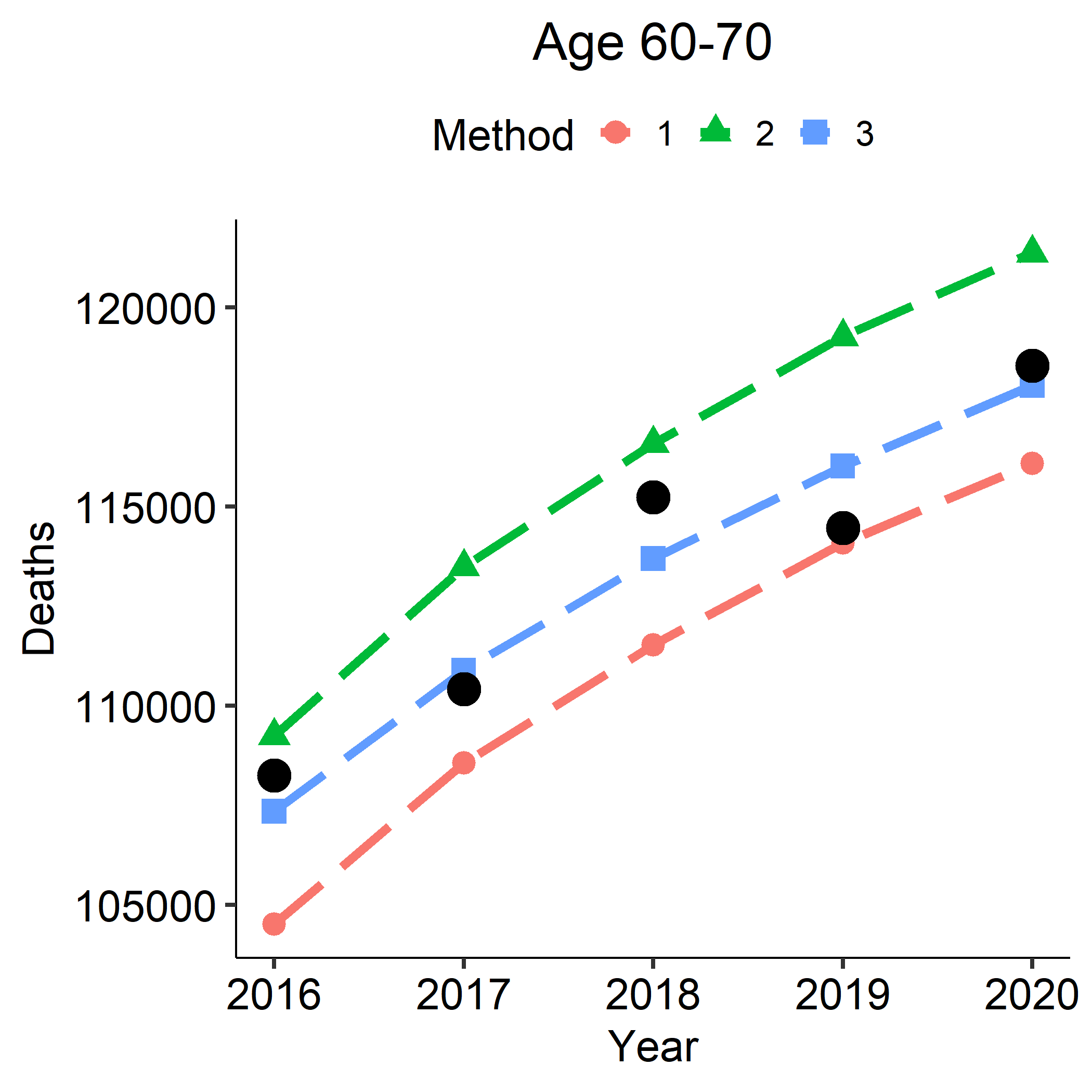}
	\includegraphics[width=0.37\linewidth]{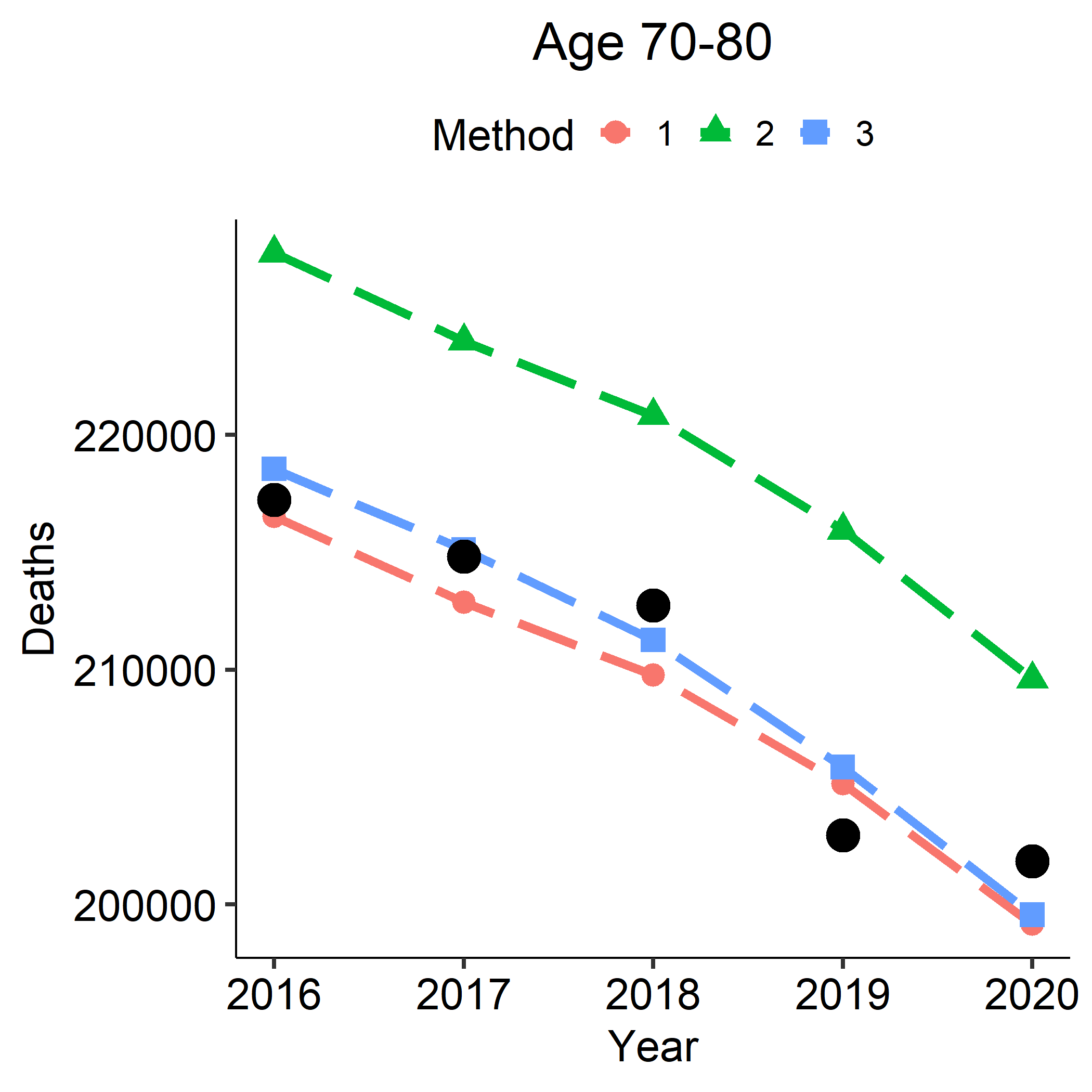}
	\includegraphics[width=0.37\linewidth]{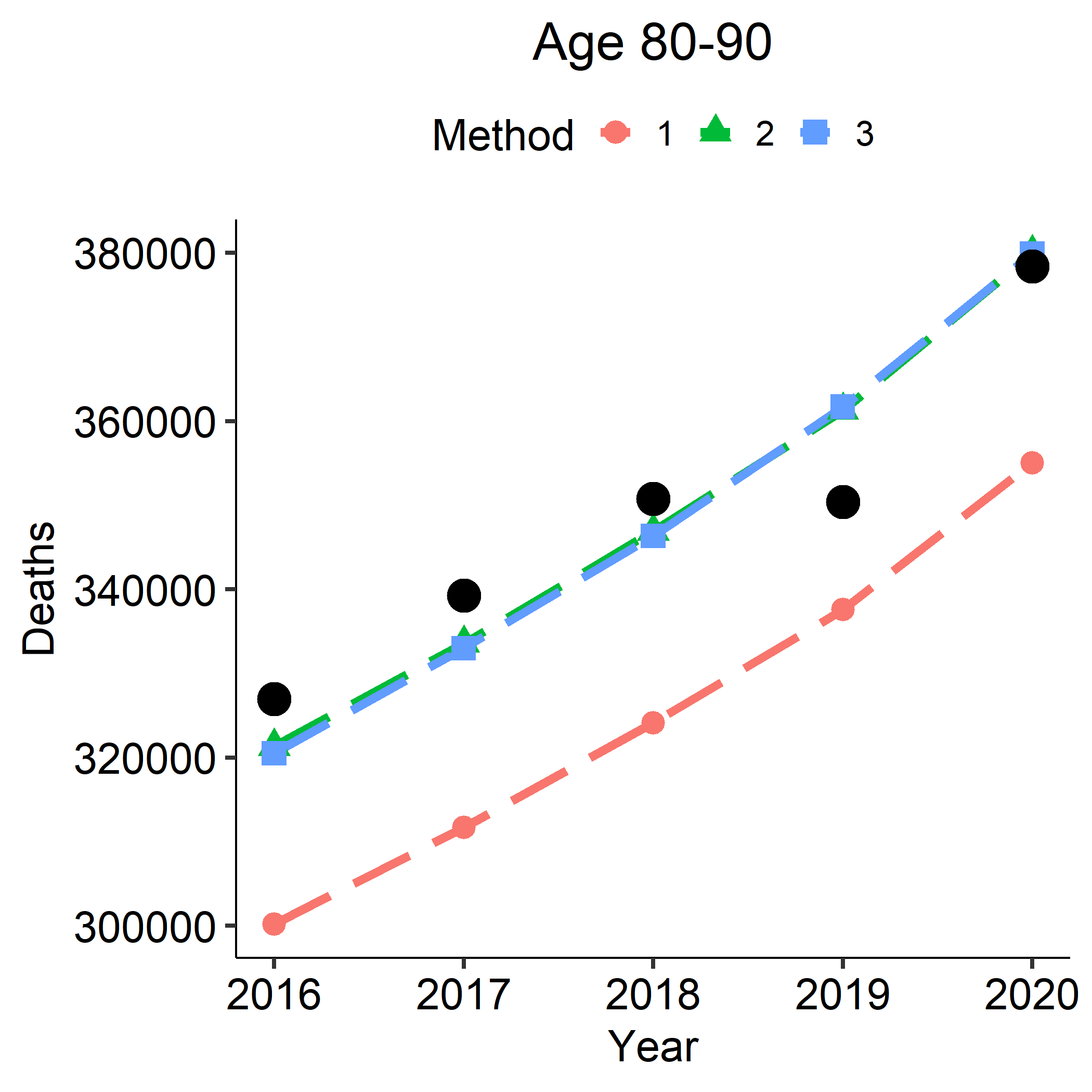}
	\includegraphics[width=0.37\linewidth]{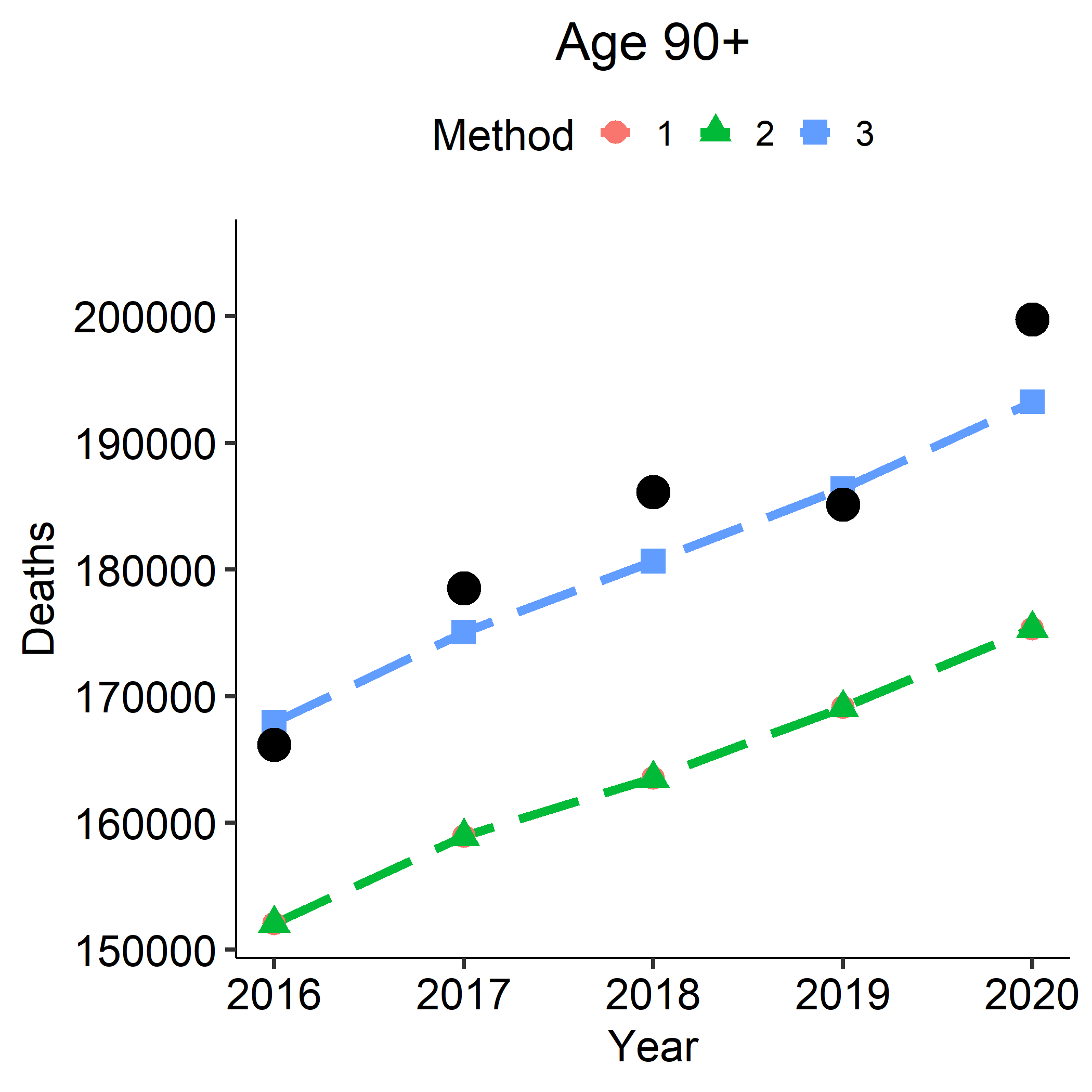}

	\caption{Expected deaths by calendar year and age group computed with the three different methods described. Realized fatalities are shown as black dots.}
	\label{fig:age}
\end{figure}

Based on these three methods we can now compare expected and observed fatalities over the last years using the same 2017/2019 life-table as basis. Note that, when looking at different years, one may more accurately also take different life tables to account for changing life expectancy. We omit this point for simplicity since we only look at five years, and changes in life expectancy over this short period were moderate \citep{Wenau:2016}. Figure \ref{fig:all} gives a first overview of the results for all age groups combined. We plot the observed death counts (black dots), and the expected counts based on the different methods are visualised as dashed lines in different colours. We can see that Method 1, which uses (\ref{eq:expected1}), clearly underestimates the expected death counts. Method 2 and Methods 3 perform equally well, which is not surprising, since we do not take an age-specific view. The latter is carried out in Figure \ref{fig:age} for all different age groups available from the data. This age-specific view shows how Methods 2 and 3 differ, and that overall Method 3 shows the better fit. We can quantify this discrepancy by calculating the root mean squared error for the different age groups, where we explicitly exclude year 2020 due to the COVID-19 pandemic. The results of this can be found in Table \ref{tab:rmse}.

Having established that Method 3 performs better than the other two over past years, we can use the expected number of fatalities computed with this method for 2020 to quantify the excess mortality during the first calendar year of the corona pandemic in Germany. Table \ref{tab:yearly} contains expected and observed mortality for all age groups in 2020, as well as the absolute and percentage variations between the two. We see from the table that, for the entire population, the age-adjusted excess mortality was in the order of 1\% in 2020. 
We stress that these results in terms of COVID-19 impact need to be interpreted with utmost care: We here focus on the methodological aspects, and defer the subject-matter discussion of the results to Section \ref{sec:discussion}. 

\begin{table}[]
	\begin{center}
		\caption{Age-specific root mean squared error for expected yearly fatalities calculated with different methods over the years 2016 to 2019. Year 2020 is excluded due to the COVID-19 pandemic. The smallest value for each age group is highlighted in bold.}
		\label{tab:table1}
		\medskip
		\begin{tabular}{l|l|l|l|l|l|l|l|l|l|} 
			{    } & \textbf{0-30} & \textbf{30-40} & \textbf{40-50} & \textbf{50-60} & \textbf{60-70} & \textbf{70-80} & \textbf{80-90} & \textbf{90+} & \textbf{Overall}\\
			\hline
			\textbf{Method 1} & 302.4 & 121.9 & {\bf 413.8} & 2221.8 & 2801.3 & 2112.7 & 24244.9 & 18374.0 & 47942.7\\
			\textbf{Method 2} & {\bf 273.0}  & 189.2 & 1052.8 & 2648.0 & 2969.6 & 10362.0 & {\bf 7038.7} & 18374.0 & 13676.7\\
			\textbf{Method 3} & 358.2 & {\bf 97.7}  & 471.6 & {\bf 1775.9} & {\bf 1207.1} & {\bf 1760.1} & 7570.3 & {\bf 3413.2 }& {\bf 13670.8} \\
		\end{tabular}
		\label{tab:rmse}
		
	\end{center}
\end{table}

\begin{table}[ht]
	\centering
	\begin{tabular}{lrrrr}
		\hline
		Age group & Expected 2020 & Observed 2020 & Absolute diff.  & Relative diff. \\ 
		\hline
		$[00,30)$ & 7471  &    7298 & -173 & -2\% \\ 
		$[30,40)$ & 6663  &    6832 &  169 & +3\% \\ 
		$[40,50)$ & 15420 &   15704 & 284  & +2\% \\ 
		$[50,60)$ & 58929 &   57606 & -1323 & -2\% \\ 
		$[60,70)$ & 118047 & 118547 & 500 & +0\% \\ 
		$[70,80)$ & 199569 & 201844 & 2275 & +1\% \\ 
		$[80,90)$ & 379917 & 378404 & -1513 & +0\% \\ 
		$[90,\infty)$ & 193238 & 199761 & 6523 & +3\% \\ 
		\hline
		Total & 979255 & 985996 & 6771 & +1\% \\ 
		\hline
	\end{tabular}
	\caption{Expected and observed yearly mortality in 2020 for each of the six age-groups, computed with Method 3.}
	
	\label{tab:yearly}
\end{table}

\section{Weekly Excess Mortality} \label{sec:exmo_weekly}
The yearly view presented in the previous section does not allow to take within-year seasonality into account for the expected deaths. We therefore now look at weekly excess mortality statements. Classical standardization approaches such as direct and indirect standardization can be used to adjust the observed values for age effects, see e.g. \citet{kitagawa1964}. We will focus on indirect standardization, but given an appropriate choice of reference population, direct standardization approaches are straightforward adaptations.

\begin{figure} [h!]
	\centering
	\includegraphics[width=\textwidth]{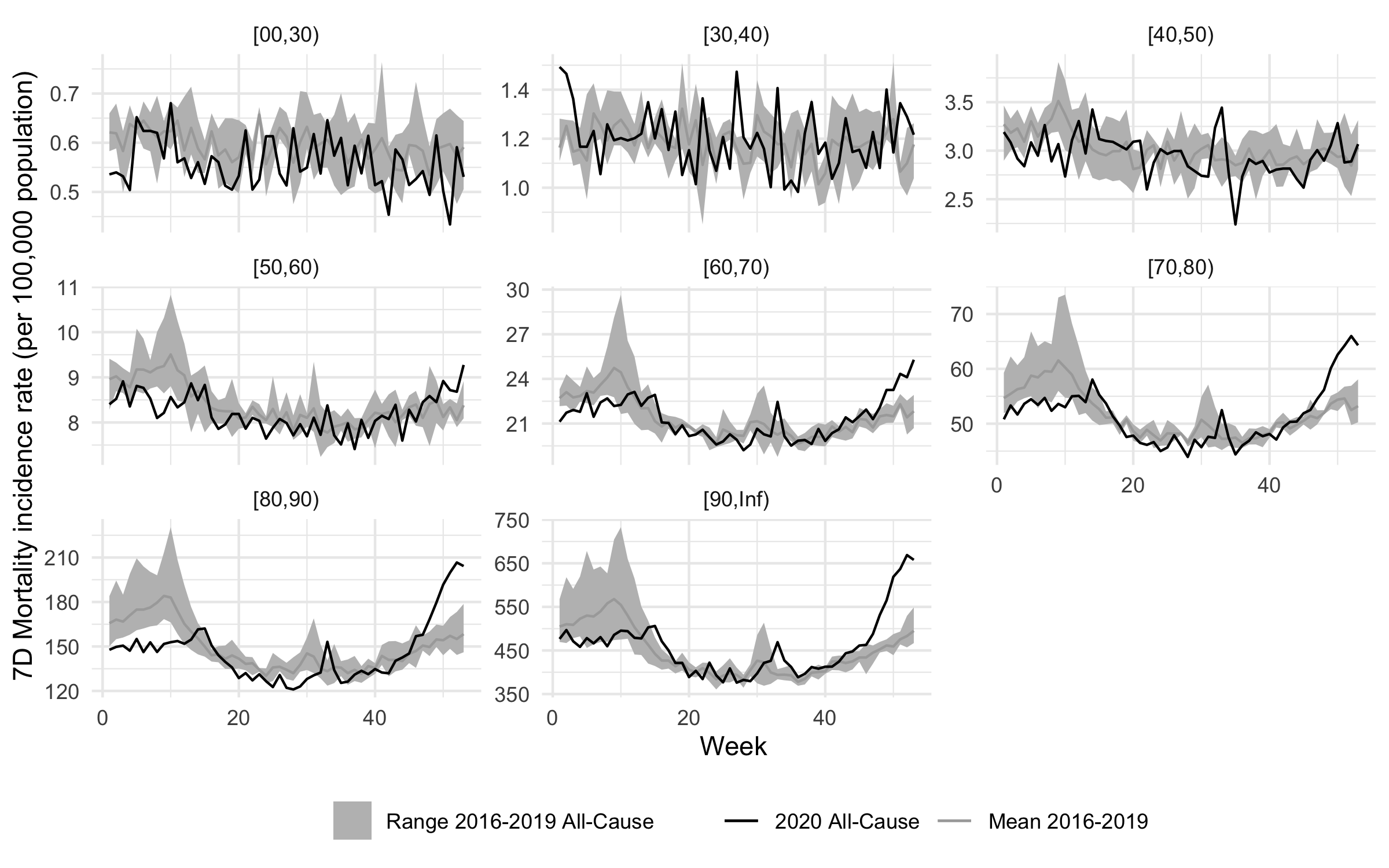}
	\caption{Weekly mortality probability estimates $\hat{q}_{t,a}$ as well as the range (min-max) of the corresponding mortality probablities of the past four years and their mean  $\overline{q}_{t,x}$.}
	\label{fig:mort_weekly_6ag}
\end{figure}

Let $q_{t,x}$ be the mortality probability specific to age $x$ and time period $t$. In what follows, the considered time period will be one International Organization for Standardization (ISO) week, but other intervals (e.g. months) are also imaginable. We estimate $q_{t,x}$ by dividing the number of observed deaths at age $x$ during time period $t$, defined as $D_{t,x}$, by the corresponding population at the beginning of the time period, i.e.\ $P_{t,x}$. To be specific, we define 
\begin{align}
	\label{eq:weekest}
	\hat{q}_{t,x}=\frac{D_{t,x}}{P_{t,x}}.
\end{align}
Since the age-stratified population is only available as a point estimate for the 31st of December of each year, we use linear interpolation to estimate $P_{t,x}$. Furthermore, the exact population of the current year, i.e.\ on December 31st, 2020, is not known at the time of analysis. We thus use a corresponding population projection: Similarly to \cite{ragnitz2021}, we use the Destatis variant G2-L2-W2\footnote{\url{https://www.destatis.de/DE/Themen/Gesellschaft-Umwelt/Bevoelkerung/Bevoelkerungsvorausberechnung/Publikationen/Downloads-Vorausberechnung/bevoelkerung-bundeslaender-2060-5124205199024.html}}. The corresponding estimates of weekly mortality probabilities (\ref{eq:weekest}) are shown in Figure 
\ref{fig:mort_weekly_6ag}. We see that in age groups $\geq 50$ years a substantial weekly excess mortality is observable from week 45 on, with more pronounced excess mortality for the elderly.

A weekly SMR-based excess mortality measure for the entire year 2020 can now be computed as follows.
Let $t$ denote a specific ISO week in 2020, i.e.\ this will serve as notational shorthand for ISO week 2020-W$t$, where $t=1,\ldots, 53$. We form the expected
age-time mortality probability for this week by computing 
the average of the mortality of the same week over the last 4 years, i.e.
$$
\overline{q}_{t,x} = \frac{1}{4} \sum_{y=2016}^{2019} \hat{q}_{y\text{-W}t,x}, \quad t=1,\ldots, 53.
$$
Because the years 2016-2019 do not have an ISO week 53, we define $y\text{-W}53$ for $y=2016,\ldots, 2019$ as $\frac{1}{2}(q_{y\text{-W}52} + q_{(y+1)\text{-W}01})$. 
The indirect standardization now computes the expected number of deaths for week $t$ as
$$
\overline{e}_{t,x} = \overline{q}_{t,x} \cdot P_{t,x}
$$
This corresponds to the expected number of deaths in week $t$ at age $x$, if the current population would have been subject to the average death probability over the past 4 years. Since fatalities are not given with exact ages but rather by age group, we indicate this by using $q_{t,A}$, $P_{t,A}$ and $e_{t,A}$, where $A$ denotes the age classes. For the available Destatis mortality data the six groups are $[00-30), [30-40), [40-50), [50-60),[60-70),[70-80),[80,\infty)$. Fig.~\ref{fig:mort_weekly_6ag} shows $\hat{q}_{t,A}$ as well as $\overline{q}_{t,x}$ for Germany. Note that the comparison for week 53 with the past year is done using the imputation scheme described above. Also note that this computation is equivalent to computing, for each reference year $y$, the expected number of deaths for the relevant week in 2020, and then taking the average of the expected deaths. In other words: by applying the mortality probabilities for the same week of the reference year $y$ to our study population (i.e.\ 2020-W$t$) and then averaging the four expected fatalities, we get:
$$
\overline{e}_{t,x} = \frac{1}{4} \sum_{y=2016}^{2019} q_{y\text{-W}t,x} \cdot P_{t,x}.
$$
One can now define the absolute excess mortality in week $t$ and age-group $A$ as $D_{t,A}-e_{t,A}$.
Instead of focusing on absolute differences, it is better in terms of interpretation to look at  relative estimates of excess mortality given by the standardized mortality ratio (SMR)
\begin{align}
	\label{eq:SMR1}
	SMR_{t,A} =   \frac{D_{t,A}}{\overline{e}_{t,A}}.
\end{align}

\begin{figure} [h!]
	\centering
	\includegraphics[width=\textwidth]{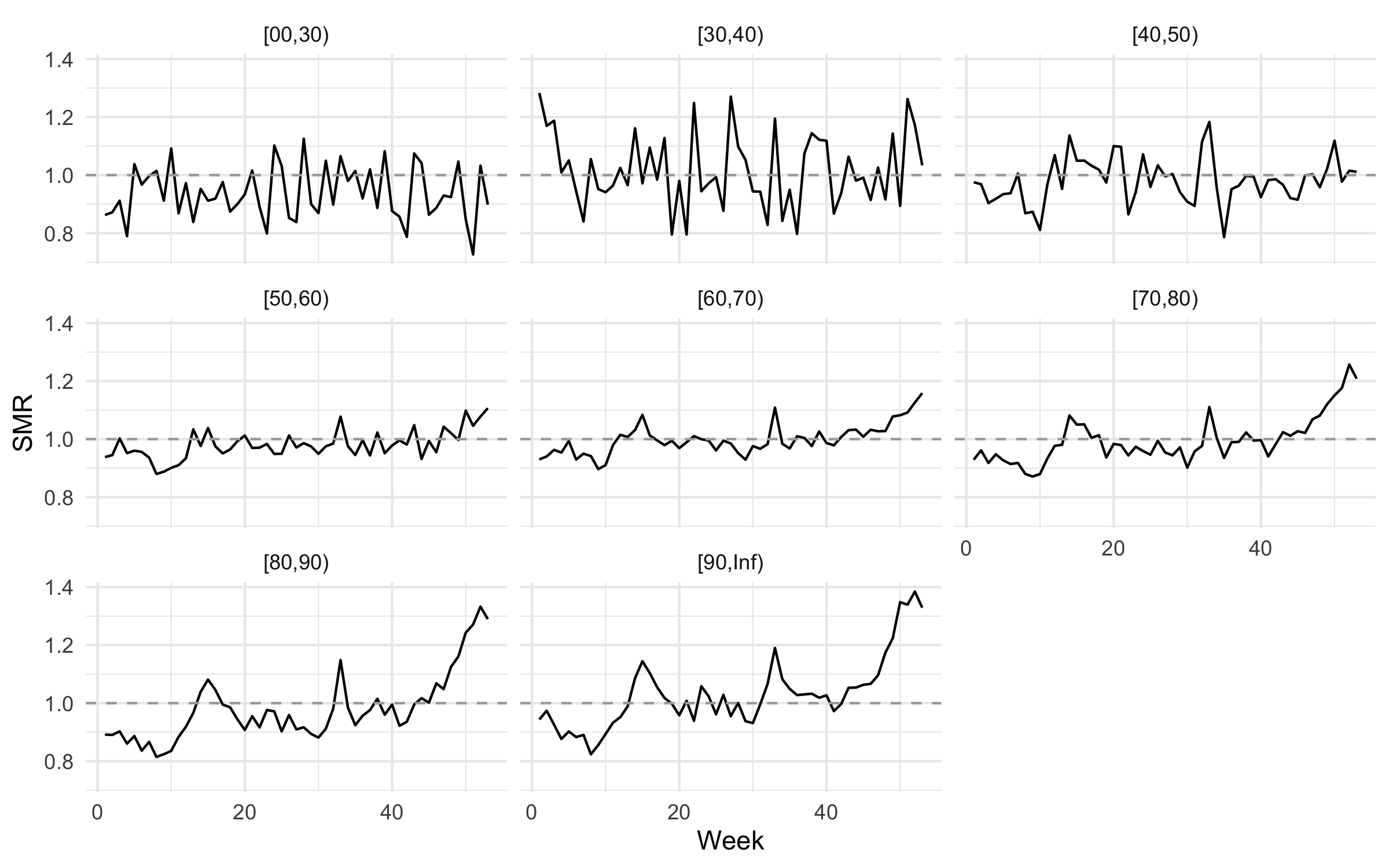}
	\caption{Weekly SMR estimates for the eight different age groups.}
	\label{fig:smr_ts}
\end{figure}

We plot the corresponding weekly estimate resulting from (\ref{eq:SMR1}) for all age groups in Figure~\ref{fig:smr_ts}. As already seen in the incidence plots, we note that in the older age groups the first approx. 10 weeks of the year had a rather low SMR, followed by a small increase consistent with the first COVID-19 wave.
Furthermore, substantial increases are then seen in in the $\geq 50$ year old age groups starting from week 45, coinciding with the 2nd wave, and reaching up to 40\% more deaths than expected in certain weeks. 

If we instead aggregate observed and expected counts per year, we could also generate yearly excess-mortality statements similar to Tab.~\ref{tab:yearly} (see e.g. \citealp{hoehle2021} for comparison).

\subsubsection*{Direct standardization}
Whereas the indirect standardization strategy, pursued above, extrapolates the average death probability from the past to the current population, an alternative is to apply the mortality probabilities from each reference year to a common standard population and then compare these numbers. This approach is, e.g., used by Statistics Austria\footnote{\url{https://www.statistik.at/web_de/presse/125475.html}} and uses the Eurostat 2013 population as common reference\footnote{\url{https://ec.europa.eu/eurostat/documents/3859598/5926869/KS-RA-13-028-EN.PDF/e713fa79-1add-44e8-b23d-5e8fa09b3f8f}}:
$$
\overline{e}_{y\text{-W}t,A}^{\text{s}} = q_{y\text{-W}t,A} \cdot P_{a}^{\text{s}},
$$
where $P_{A}^{\text{s}}$ denotes the size of the standard population in age-group $A$ and the expected number of deaths for the week $t$ in 2020 is given by $e_{t,A}^s = \sum_{y=2016}^{2019} \overline{e}_{y\text{-W}t,A}^{\text{s}}/4$.

\section{Discussion}\label{sec:discussion}

The COVID-19 pandemic posed numerous challenges to scientists. One of those challenges lies in estimating the number of casualties brought upon by the pandemic. To tackle this issue, we pursued an approach based on comparing observed all-cause mortality in 2020 with the number of fatalities that would have been expected in the same year without the advent of COVID-19. Building on existing methodology, we proposed two simple ways of computing expected mortality at the yearly and at the weekly level. We then put those method to work to obtain estimates for excess mortality in 2020 in Germany. The two approaches yield similar results at the aggregate level, and highlight how 2020 was characterized by an overall excess mortality of approximately 1\%. The light excess mortality was apparently driven by a spike in fatalities related to COVID-19 at the end of the year in older age groups.

Interpreting COVID-19 mortality has become a  politically sensitive issue, where the same underlying data are used to either enhance or downplay the consequences of COVID-19 infections. We therefore stress that our interests are methodological, and that the presented results are restricted to the calendar year 2020 for Germany as a whole. Altogether, the mild mortality in the older age groups during the first weeks (e.g.\ due to a mild influenza season) balanced the excess in the higher age groups which came later in the year. Clearly noticeable is the second wave during Nov-Dec 2020, which also continued in the early months of 2021. To better account for such seasonality, excess mortality computations for influenza are often pursued by season instead of calendar year, i.e.\ in the northern hemisphere for the period from July in Year $X$ to June in Year $X+1$ \citep{nielsen_etal2011}. Similarly, the impact of COVID-19 cases and fatalities was not only temporally, but also spatially  heterogeneous, with strong peaks in Dec 2020 in the federal states of Saxony, Brandenburg and Thuringia~\citep{hoehle2021}. Hence, using mortality aggregates over periods and regions only provides a partial picture of the impact of COVID-19. Furthermore, the mortality figures observed in 2020 naturally incorporate the effects of all types of pandemic management consequences, which include changes in the behavior of the population (voluntary or due to interventions). Disentangling the complex effects of all-cause mortality and the COVID-19 pandemic is a delicate matter, which takes experts in several disciplines (demographers, statisticians, epidemiologists) to solve. Timely analysis of all-cause mortality data is just one building block of this process; Nevertheless, the pandemic has shown the need to do this in near real-time based on sound data while adjusting for age structure. 

Our analysis was motivated by the fact that many of the methods that have been applied to tackle this issue so far fail to take the changing age structure of the population into account. This can lead to biased results, and especially so for the rapidly aging developed countries. In the case of Germany, for example, the absolute number of people aged 80 or more increased by approximately 20\% from 2016 to 2020. Such a remarkable increase will naturally have an effect on overall mortality, and as such direct comparisons in the number of casualties across different years will lead to significant overestimation of the excess mortality. Our approaches are instead robust to such changes in population structure, and can be used regardless of the demographic context. Note that, for both of our approaches, it would also be possible to obtain confidence intervals through imposing simple distributional assumptions. The same methodologies could be used to pursue a similar analysis for any country in which mortality data and a mortality table are available, for any given year. A natural use for the proposed methodology would also be to assess the overall damages caused by the pandemic when it will be finally considered a thing of the past. 
All in all, we hope the proposed methods will help shedding light on the issue of computing the expected number of fatalities, and in the assessment of potential excess mortality.

\bibliography{library}
\end{document}